\documentclass[pra,letterpaper,aps,10pt,superscriptaddress,twocolumn,floatfix,showpacs]{revtex4-1}
\usepackage{graphicx}
\usepackage{amsmath}
\usepackage{amsfonts}
\usepackage{amssymb}
\usepackage{epsfig}
\usepackage{epstopdf}
\DeclareGraphicsExtensions{.pdf,.eps,.png,.jpg,.mps}
\usepackage[pdftex]{color}
\usepackage{amsmath,graphicx,amssymb,braket,xcolor,subfigure,upgreek}
\usepackage[colorlinks, linkcolor=blue, citecolor=blue, urlcolor=blue, breaklinks=true]{hyperref}
\usepackage{microtype}
\usepackage{bbm}
\usepackage{color}

\begin{document}

\title{Ramsey interferometry of Rydberg ensembles inside microwave cavities}
\author{Christian Sommer and Claudiu Genes}
\affiliation{Max Planck Institute for the Science of Light, Staudtstra{\ss}e 2,
D-91058 Erlangen, Germany}
\date{\today}

\begin{abstract}
We study ensembles of Rydberg atoms in a confined electromagnetic environment such as provided by a microwave cavity. The competition between standard free space Ising type and cavity-mediated interactions leads to the emergence of different regimes where the particle-particle couplings range from the typical van der Waals $r^{-6}$ behavior to $r^{-3}$ and to $r$-independence. We apply a Ramsey spectroscopic technique to map the two-body interactions into a characteristic signal such as intensity and contrast decay curves. As opposed to previous treatments requiring high-densities for considerable contrast and phase decay~\cite{Takei2016direct,Sommer2016time}, the cavity scenario can exhibit similar behavior at much lower densities.
\end{abstract}


\maketitle

\section{Introduction} \label{intro}
Long-range interactions in many-body systems have recently become of central interest~\cite{Britton2012Engineered,Martin2013AQuantum,Yan2013Observation,Richerme2014Non}. A widely experimentally investigated platform employs interacting Rydberg atom ensembles where the evolution is governed by an Ising-type Hamiltonian with couplings going beyond the nearest neighbor~\cite{Schauss2015Crystallization,Zeiher2016Many,Takei2016direct}. Such investigations are mainly geared towards describing regimes of strong quantum correlations and towards quantum simulations~\cite{Bloch2008many,Labuhn2016Tun,Bernien2017Probing,Zhang2017Observation}. In standard scenarios (such as free space), the Ising-type Hamiltonian comes from an effective $r^{-6}$ van der Waals interaction between excited levels (optically addressable from the ground state) of neighboring atoms. The interaction is an effective one and stems from a perturbative treatment of the near-field dipole-dipole coupling scaling as $r^{-3}$ between a manifold of Rydberg states in the vicinity of the level of interest (at frequencies between $100$ GHz to $10$ THz). For such fastly decaying potentials, high density samples are usually employed to allow the emergence of strongly correlated many body dynamics~\cite{Takei2016direct}.

\begin{figure}[b]
\includegraphics[width=0.65\columnwidth]{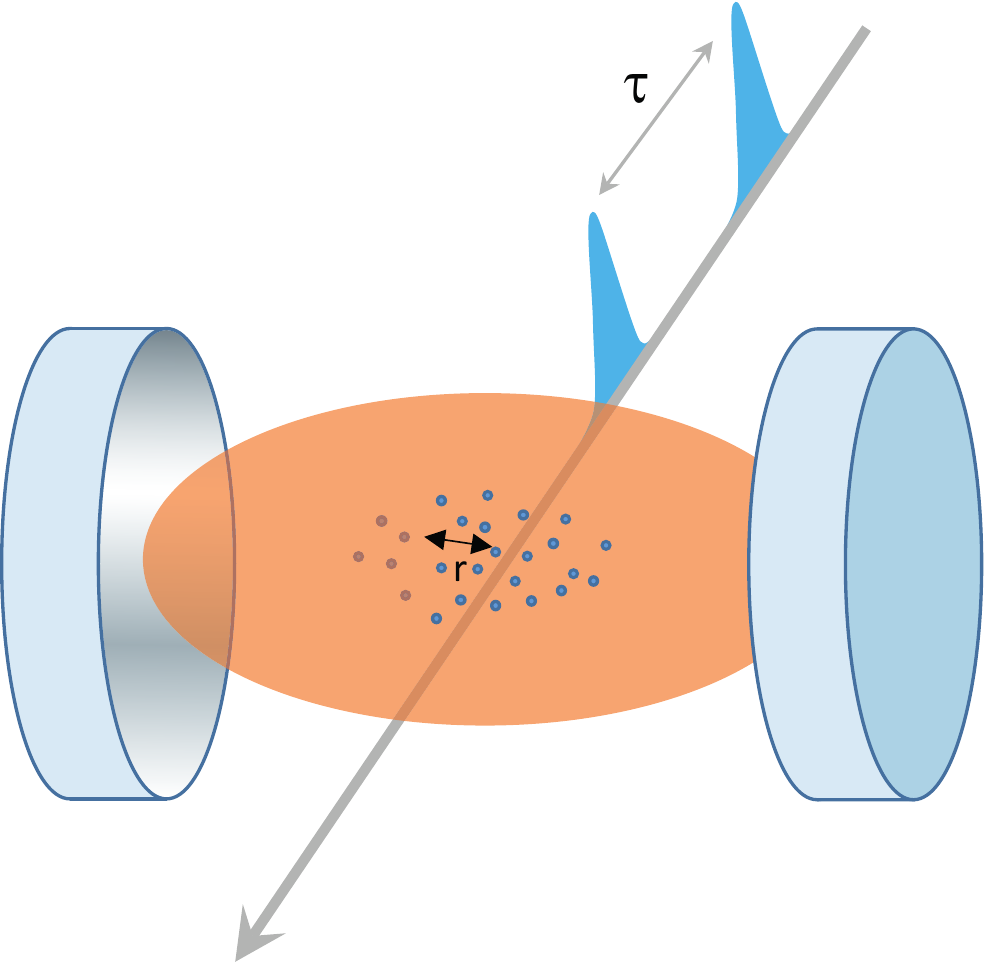}
\caption{\emph{Intracavity Ramsey interferometry}. Sketch of the time-delayed two pulses Ramsey spectroscopic process on a dilute ensemble of Rydberg atoms located within a microwave cavity wavelength.}
\label{fig1}
\end{figure}

In this paper we propose to replace the free space mechanism of dipole-dipole coupling with a microwave cavity mediated interaction. In a perturbative regime, one expects that, for large enough distances, the cavity-mediated interaction would be dominant and an all-to-all distance-independent coupling would occur similarly as has been obtained in \cite{Pupillo2010Strongly,Britton2012Engineered,Zeiher2016Many}. To this end we derive particle-particle interactions via the microwave cavity modes and analyze the scaling from small to large distances. We find tunable regimes describing all to all interaction at long distances \cite{Petrosyan2008Quantum} followed by a $r^{-3}$ scaling in the intermediate range transiting in a counterintuitive manner into a $r^{-6}$ van der Waals scaling for short internuclear distances. Extending the derived results to large ensembles, we show that the dynamics of such a system can be read out by spectroscopic methods. In particular we employ a Ramsey interferometry sequence [see Fig.~\ref{fig1}] where two identical pulses map the coherence of the atoms into population in the excited state as a function of the delay time $\tau$ between the pulses \cite{Ramsey1950AMolecular}. Here, the particular features of the interaction can result in characteristic outcomes of the Ramsey signal. Using procedures previously explored in~\cite{ Sommer2016time} we find analytical solutions for the Ramsey contrast in the large particle number limit.\\
The paper is structured as follows. Sec.~II introduces the model for a pair of 4 level atoms interacting via the free space and cavity field modes. In Sec.~III we present a perturbative derivation of the cavity-mediated interactions and analyze the resulting regimes. In Sec.~IV we apply this model to a typical Ramsey interferometry setup. We discuss possible experimental feasibility in Sec.~V.

\section{Model} \label{Model}
\begin{figure}[b]
\includegraphics[width=0.65\columnwidth]{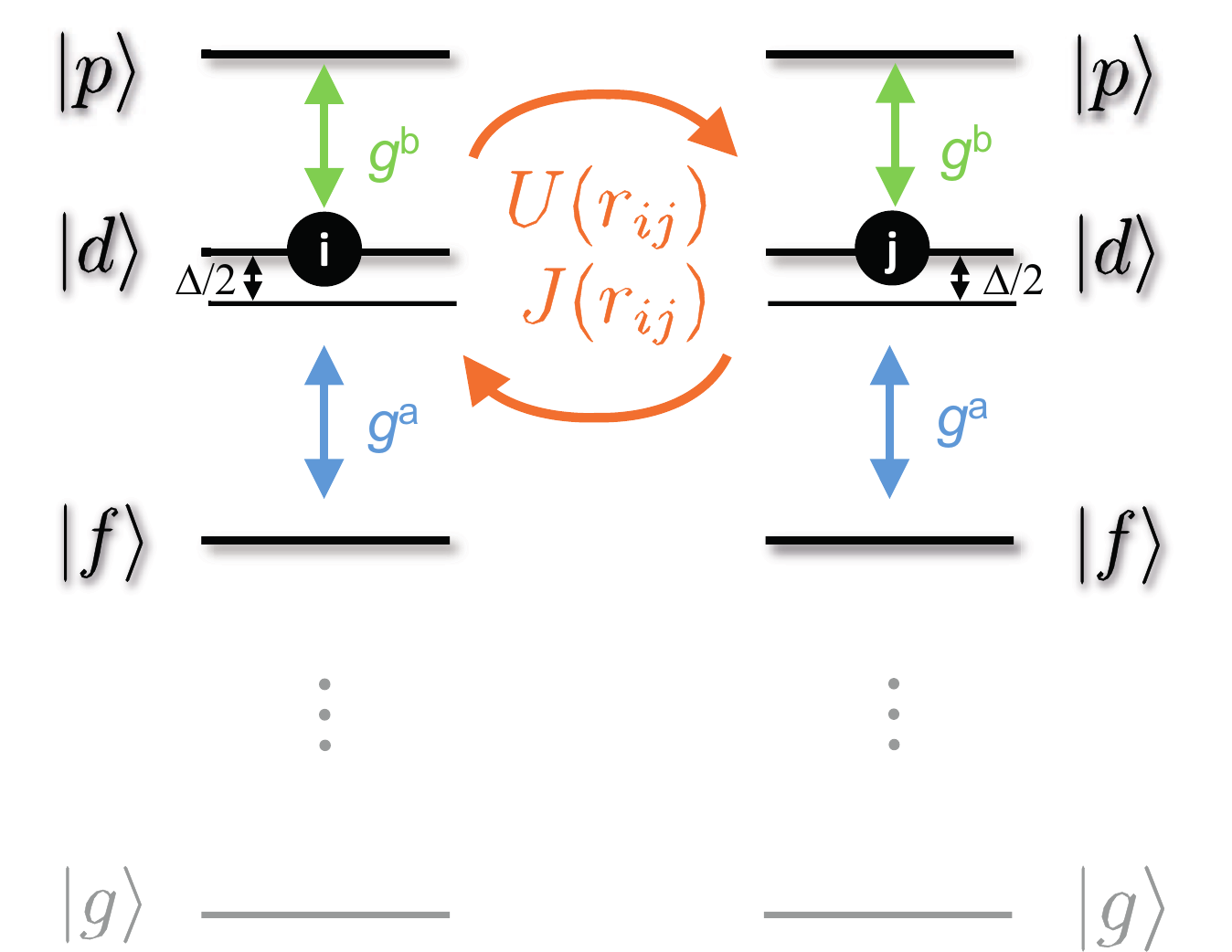}
\caption{\emph{Two atom system}. a) The level scheme involves the Rydberg states $\ket{f}$, $\ket{d}$ and $\ket{p}$ and an energetically distant ground state $\ket{g}$. Free space and cavity coupling rates are highlighted in the figure.}
\label{fig2}
\end{figure}

We consider an ensemble of Rydberg atoms inside a microwave cavity and subjected to a two-pulse time-domain Ramsey interferometric scheme as depicted in Fig.~\ref{fig1}. The relevant considered internal structure [see Fig.~\ref{fig2}] of each atom is given by ground level $g$ and excited level $d$ typically reachable via a two-photon optical transition; in addition two adjacent states $f$ and $p$ are considered, lying in the neighborhood of $d$ and accessible from it via microwave photons. The cavity mode, at frequency $\omega$ (in the microwave range) can mediate transitions between levels $f$ and $d$ and $d$ and $p$, respectively (Rydberg states are typically separated by frequencies on the order of $100$ GHz to $10$ THz). We assume a preparation stage where an excitation scheme is employed to selectively drive the atoms from the ground state $g$ directly to the $d$ state via a two-photon process in the optical domain. Afterwards, we are solely interested in the dynamics within the Rydberg $d,p,f$ manifold (assuming that the relevant dynamics is on a much faster timescale than the lifetime of the $p,d,f$ states). The free Hamiltonian for a given particle $i$ can be expressed as
\begin{equation}
\label{Eq.M1}
H^0_i=\omega_d \ket {d_i} \bra {d_i}+\omega_p \ket {p_i} \bra {p_i}+\omega \hat{a}^\dagger \hat{a},
\end{equation}
where level $f$ has been set at the zero energy level such that levels $d$ and $p$ have energies $\omega_{d,p}$ (with $\hbar$ set to unity). The operators  $\hat{a}$, $\hat{a}^{\dagger}$ are the annihilation and creation operators with respect to the cavity mode of frequency $\omega$. The transition dipoles between $f \leftrightarrow d$ and $d \leftrightarrow p$ are defined by $\mu^{a}$ and $\mu^{b}$, respectively. The direct, free-space mediated interaction between two atoms indexed by $i$ and $j$ is
\begin{eqnarray}
\label{Eq.M2}
\nonumber
H^{F}_{ij} = &U_{ij}&\left( \ket {f_i} \bra {d_i} \otimes \ket {p_j} \bra {d_j} + \ket {p_i} \bra {d_i} \otimes \ket {f_j} \bra {d_j}\right)+\\
   + &J_{ij}& \ket{f_i} \bra{d_i} \otimes \ket{d_j} \bra{f_j}+ h.c. ,\\ \nonumber
\end{eqnarray}
where we have neglected the anti-resonant terms that couple $d_i d_j \leftrightarrow f_i f_j$ and $d_i d_j \leftrightarrow p_i p_j$ due to their large detuning of $2\omega_d$ and $2\omega_{p}$. Additionally we have also neglected terms coupling $d_i p_j \leftrightarrow p_i d_j$ since they would require the initial presence of a photon (we consider zero temperature environments).
The terms $U_{ij}$ and $J_{ij}$ mediate dipole-dipole interactions and generally have a very complex dependence on the angle $\theta$ (between the dipole direction and the interparticle axis $\mathbf{r}_{ij}$) characterizing the anisotropy of interaction. We restrict our treatment to the following expressions:
\begin{eqnarray}
\label{Eq.M3}
U_{ij} &=& \frac{\mu^{a}\mu^{b}}{4\pi\epsilon_0 r^{3}_{ij}}\left[1 - 3\cos^2 (\theta)\right] \\
\label{Eq.M3a}
J_{ij} &=& \frac{\left(\mu^{a}\right)^{2}}{4\pi\epsilon_0 r^{3}_{ij}} \left[1-3\cos^2(\theta)\right],
\end{eqnarray}
while pointing out that different functions can be obtained by addressing suitable states of the Rydberg manifolds and/or manipulating the cavity mode polarization~\cite{Reinhard2007Level}.

The cavity-atom couplings are standard Jaynes-Cummings interactions:
\begin{equation}
\label{Eq.M4}
H^{JC}_{i}=g^a_i \hat{a}^{\dagger} \ket {f_i} \bra {d_i}+g^b_i \hat{a}^{\dagger} \ket {d_i} \bra {p_i} +h.c.,
\end{equation}
where $g^{a,b}_i = \mu^{a,b} \sqrt{\omega/2\epsilon_0 V}\Phi(\mathbf{x}_i)$ give the coupling between the cavity field and the atomic states. The cavity mode function $\Phi(\mathbf{x})$ is evaluated at the position of the atom $i$ and $V = \int |\Phi(\mathbf{x})|^2 d^{3}\mathbf{x}$ determines the mode volume of the electromagnetic field. To simplify our notation the full Hamiltonian is expressed as $H = H^{0} + H^{1}$ where $H^{0} = \sum_{i} H^{0}_{i}$ and $H^{1} = \sum_{i} H^{JC}_{i} + \sum_{i < j} H^{F}_{ij} $.\\
Also we assume that the dipole moments of the atoms point along the same orientation which can be obtained by bringing the atoms to the same magnetic sublevel of the electronic excited states.

\section{Cavity mediated effective interactions}
\label{Cavity}
Let us analyze the role of the cavity in mediating interactions between pairs of atoms $i,j$ separated by distance $r$ within the ensemble.  In the two excitation subspace, the full Hamiltonian can be written in matrix form in the two particle basis $\ket{df1}$, $\ket {fd1}$, $\ket {ff2}$, $\ket {pf0}$, $\ket {fp0}$, $\ket {dd0}$ as presented in Appendix A. We define the detunings $\delta = \omega_{d}-\omega$ between the cavity resonance $\omega$ and the atomic transition $\ket{f} \leftrightarrow \ket{d}$ and $\Delta = 2\omega_{d}-\omega_{p}$ as the F\"orster detuning between the two particle states $\ket{dd}$ and $1/\sqrt{2}(\ket{pf}+\ket{fp})$. Our approach is closely related to investigations of van der Waals interactions of the ground states of two-level systems beyond the Jaynes-Cummings approximation in planar cavities carried out in Ref.~\cite{Donaire2017Dipole}.

\subsection{Results in the perturbative regime}

In the regime of sufficiently large detunings $\delta, \Delta \gg g^{a},\; g^{b},\; U,\; J$ we can simplify the system even further so that effectively only the $\ket{dd0}$ state is considered. By applying perturbation theory we acquire an effective interaction between two atoms in $|d\rangle$ states. The interaction can be obtained from the energy shifts up to fourth order which is necessary to acquire all the coupling terms for the different distance regimes
\begin{widetext}
\begin{eqnarray}
\label{Eq.C3}
\Delta E_{1} &=& \langle dd0| H^{1}|dd0\rangle = 0,\\
\label{Eq.C4}
\Delta E_{2} &=& \sum_{n \neq dd0} \frac{\langle dd0| H^{1}|n\rangle \langle n| H^{1}|dd0\rangle}{E_{dd0} - E_{n}}
= \frac{\left( g^{a}_{1}\right)^{2}}{\delta} + \frac{\left( g^{a}_{2}\right)^{2}}{\delta} + \frac{2U^{2}}{\Delta},
\label{Eq.C5}
\end{eqnarray}
\end{widetext}
and already reveal the free space direct van der Waals interaction in second order perturbation theory. Novel interaction terms are obtained from the third and fourth order calculation
\begin{widetext}
\begin{eqnarray}
\Delta E_{3} &=& \sum_{n,l \neq dd0} \frac{\langle dd0| H^{1}|n\rangle \langle n| H^{1}|l\rangle\langle l|H_{1}|dd0\rangle}{(E_{dd0}-E_{n})(E_{dd0}-E_{l})}
= \frac{2(g^{a}_{1}g^{b}_{2}+g^{a}_{2}g^{b}_{1})U}{\Delta\delta} + \frac{2g^{a}_{1}g^{a}_{2}J}{\delta^{2}},\\
\label{Eq.C6}
\Delta E_{4} &=& \sum_{n,l,k \neq dd0} \frac{\langle dd0| H^{1}|n\rangle \langle n| H^{1}|l\rangle\langle l|H^{1}|k\rangle\langle k|H^{1}|dd0\rangle}{(E_{dd0}-E_{n})(E_{dd0}-E_{l})(E_{dd0}-E_{k})}-\sum_{n,l \neq dd0} \frac{|\langle dd0| H^{1}|n\rangle|^2 |\langle dd0| H^{1}|l\rangle |^2}{(E_{dd0}-E_{l})^{2}(E_{dd0}-E_{n})}\\\nonumber
&=& \frac{(g^{a}_{1}g^{b}_{2})^{2}+(g^{a}_{2}g^{b}_{1})^{2}}{\delta^2\Delta}+ \frac{2(g^{a}_{1})^{2}(g^{a}_{2})^{2}}{\delta^3} - \frac{2U^{2}\left((g^{a}_{1})^{2} + (g^{a}_{2})^{2}\right)}{\Delta\delta} \left(\frac{1}{\Delta}+\frac{1}{\delta}\right)+ \frac{U^{2}\left((g^{b}_{1})^{2}+(g^{b}_{2})^{2}\right)}{\Delta^{2}\delta} - \frac{4U^{4}}{\Delta^{3}}\\\nonumber
& & - \frac{(g^{a}_{1})^{4}}{\delta^{3}} - \frac{(g^{a}_{2})^{4}}{\delta^{3}} + \frac{2JU(g^{a}_{1}g^{b}_{1} + g^{a}_{2}g^{b}_{2}) }{\delta^{2}\Delta} + \frac{\left((g^{a}_{1})^{2} + (g^{a}_{2})^{2} \right)J^{2}}{\delta^{3}}.
\end{eqnarray}
\end{widetext}

\begin{figure}[t]
\includegraphics[width=1.0\columnwidth]{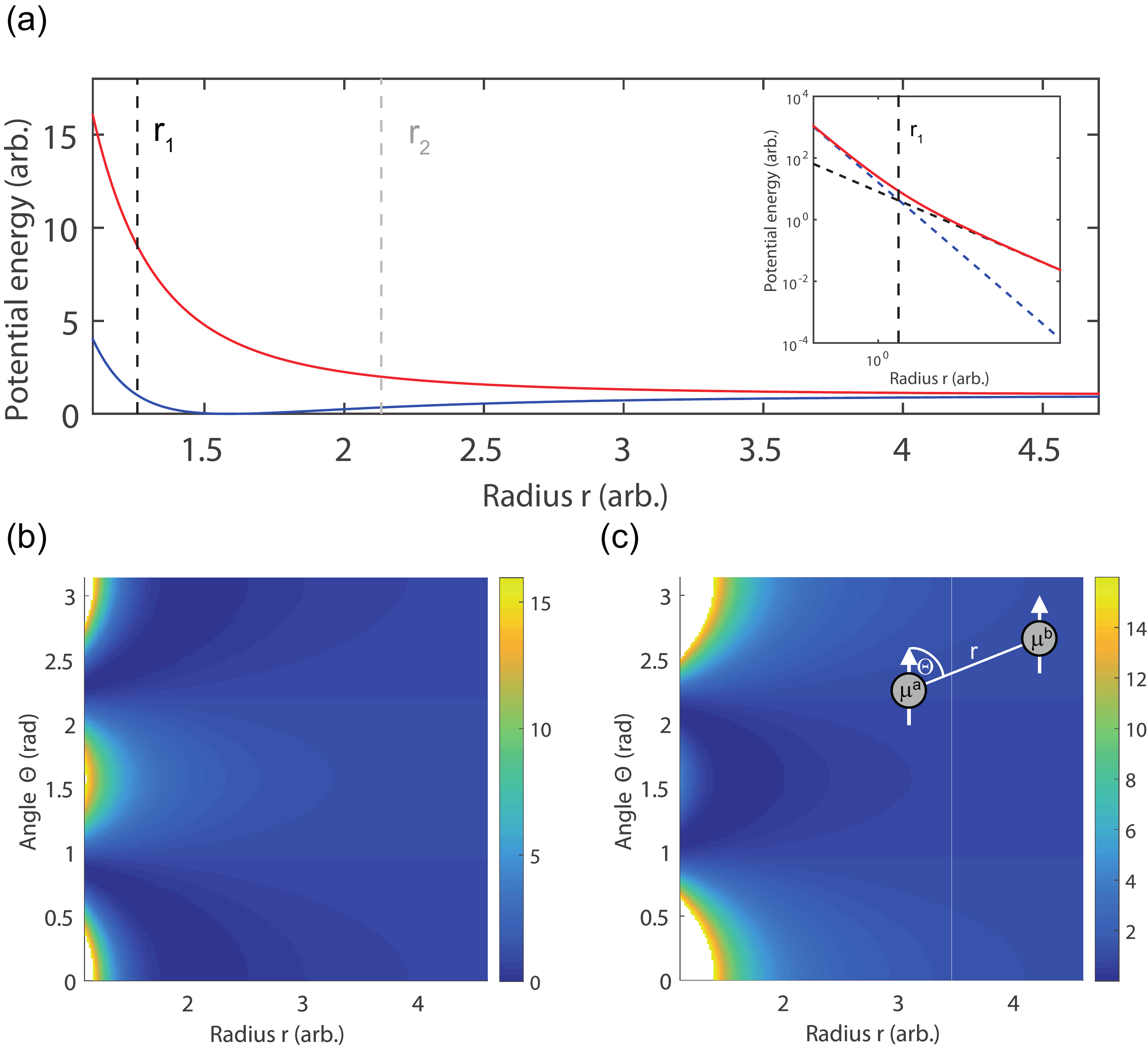}
\caption{\emph{Effective potential}. (a) The effective van der Waals in the presence of the cavity environment for dipole orientation described by $\theta = \pi/2$. The radii $r_{1}$ and $r_{2}$ roughly indicate where the interaction changes from $\propto r^{-6}$ to $\propto r^{-3}$ and $r^{-3}$ to $r$ independence, respectively. The inset shows a log-log plot of the potential in the case that $C_{0} = 0$ and $C_{3,6}$ are both positive and reveals more clearly the change from $\propto r^{-6}$ to $\propto r^{-3}$ dependence indicated by the dashed curves where the vertical line locates $r_{1}$. (b) The potential is plotted for different angles and radii. (c) In the case of a detuning or coupling constants that allow for a different sign for the terms proportional to $r^{-6}$ and $r^{-3}$ we obtain a minimum at $\theta = \pi/2$ which results in a weak binding potential for two dimensional arrangements. The parameters chosen here for the potential are $C_{6} = 16C_{0}$ and $C_{3} = 8C_{0}$ and  $C_{0} = 1$. }
\label{fig3}
\end{figure}

\noindent Here, $\Delta E^{\mathrm{tot}} = \sum_{i=1}^{4}\Delta E_{i}$ is the total energy shift of the $|dd0\rangle$ state up to fourth order. Besides the terms resulting from ac-Stark shifts up to fourth order $\Delta E^{\mathrm{tot}}_{i} = (g^{a}_{i})^{2}/\delta- (g^{a}_{i})^{4}/\delta^{3}$, the dominant two particle interaction terms are given by
\begin{eqnarray}
\label{Eq.C7}
\Delta E^{\mathrm{tot}}_{ij} &\approx& \frac{U_{ij}^{2}}{\Delta} + \frac{2U_{ij}g^{a}_{i}g^{b}_{j}}{\Delta\delta} + \frac{J_{ij}g^{a}_{i}g^{a}_{j}}{\delta^{2}}\\\nonumber
& & + \frac{(g^{a}_{i}g^{b}_{j})^{2}}{\Delta\delta^2} + \frac{(g^{a}_{i}g^{a}_{j})^{2}}{\delta^3}.
\end{eqnarray}
This expression is found by taking only terms with $O(e^4)$ into account, where $e$ is the electric charge of an electron \cite{Holstein2001The}.
The first term in Eq.~\ref{Eq.C7} which is describing the free space van der Waals interaction dominates at short internuclear distances while the second and third term contribute strongly in the intermediate regime defined by the relation $U_{ij} \approx (2g^{a}_{i}g^{b}_{j}/\delta)(1+\Delta/2\delta)$. The last two terms govern the dynamics in the long distance regime where the all to all interaction mediated by the cavity field is dominant \cite{Petrosyan2008Quantum}.\\
Also it can easily be shown that the expression in Eq.~\ref{Eq.C7} is the general solution for the interaction between $N$ atoms in the $\ket{dd\dots d0}$ state residing in the cavity field.

\subsection{Analysis of emergent $r$-scaling regimes}

The energy shifts derived above can be casted into an effective Hamiltonian
\begin{eqnarray}
\label{Eq.C8}
H^{\mathrm{eff}} = \sum_i \tilde{\omega}_{d,i}\ket{d_i}\bra{d_i} + \sum_{i \neq j}\frac{\tilde{U}_{ij}}{2}\ket{d_i d_j}\bra{d_i d_j},
\end{eqnarray}
where $\tilde{\omega}_{d,i} = \omega_{d} + (g^{a}_{i})^2/\delta - (g^{a}_{i})^{4}/\delta^3$ and
\begin{eqnarray}
\label{Eq.C8a}
\nonumber
\frac{\tilde{U}_{ij}}{2} &=& \frac{U_{ij}^{2}}{\Delta} + \frac{U_{ij}\left(g^{a}_{i}g^{b}_{j} + g^{b}_{i}g^{a}_{j}\right)}{\Delta\delta} + \frac{\left(g^{a}_{i}g^{b}_{j}\right)^{2}+\left(g^{b}_{i}g^{a}_{j}\right)^{2}}{2\Delta\delta^2}\\
& & + \frac{J_{ij}(g^{a}_{i}g^{a}_{j})}{\delta^2} + \frac{(g^{a}_{i}g^{a}_{j})^{2}}{\delta^{3}}.
\end{eqnarray}
Features of the potential are illustrated in Fig.~\ref{fig3}. Here, the anisotropy of the dipole dipole interaction and by choosing the right values and signs for the detuning $\Delta$ and $\delta$ allows for the emergence of a weak binding potential at a specific spatial orientation (see Fig.~\ref{fig3}b,c), which would not be possible for a free space van der Waals interaction.\\
In the following derivations we will apply the assumption that $\mu^{a,b} = \mu$ are of equal magnitude which results in $g^{a,b} = g$.
From Eq.~\ref{Eq.C8a} we understand that the interaction between two atoms in the $\ket{d}$ state can be rewritten as $\tilde{U}(r) = C_{0} + C_{3}/r^{3} + C_{6}/r^{6}$, where
\begin{eqnarray}
C_{0} &=&  \frac{2g^4}{\delta^2}\left(\frac{1}{\Delta} + \frac{1}{\delta}\right),\\
C_{3} &=& \frac{2\mu^2 g^2}{4\pi\epsilon_{0}\delta}\left(\frac{2}{\Delta} + \frac{1}{\delta} \right)\\
C_{6} &=& \frac{2\mu^{4}}{(4\pi\epsilon_{0})^{2}\Delta}
\end{eqnarray}
For simplicity we ignore from now on the anisotropy of the potential which is examplified in Fig.~\ref{fig3}a,b.
For the ongoing discussion we cast the potential in the form
\begin{eqnarray}
\label{Eq.C8b}
\nonumber
\tilde{U}(r) &=& C_{6}\left[ \frac{1}{r^6} + \left(1 + \frac{\Delta}{2\delta}\right) \frac{\mathrm{sgn}(\delta)}{R^{3}r^{3}} +\frac{1}{4}\left(1 + \frac{\Delta}{\delta} \right)\frac{1}{R^{6}} \right],\\
\end{eqnarray}
where $R$ is the constant effective cavity van der Waals radius which is defined by $R^3 = |\delta|V/(4\pi\omega) $.
The different terms of the potential $\tilde{U}(r)$ have different ranges of dominance over the other terms and therefore can play prominent roles at different densities of the atomic ensemble.\\
The validity of the formalism in Eq.~\ref{Eq.C8} is given from $r \gg r_0= \sqrt{2}\sqrt[3]{\mu^2/(4\pi\epsilon_0\Delta)}$ which originates from the relation $\Delta/2 \gg \sqrt{2}U$ and is ranging to $r \approx L$. Within this range we find different regions where the interaction shows a different characteristic dependence on the distance $r$. For example, for $r \gg r_0$ and
\begin{equation}
r < r_1 = \frac{1}{\sqrt[3]{\left|1+\frac{\Delta}{2\delta}\right|}}R,
\end{equation}
the interaction is of van der Waals character with $\tilde{U} \propto r^{-6}$ and close to identical to the free space van der Waals interaction between atoms in the $\ket{d}$ state. Here, the definition of $r_{1}$ is derived from the equation $\left|C_{3}/r^{3}\right| = \left|C_{6}/r^{6}\right|$. The behavior of the interaction is changing in the range $r > r_1$ and
\begin{equation}
r < r_2 =\frac{R}{\sqrt[3]{|1+\Delta/(2\delta)|\sqrt{1+\frac{|1+\Delta/\delta|}{(1+\Delta/(2\delta))^2}}-\frac{(1+\Delta/(2\delta))}{\mathrm{sgn}(\delta)}}},
\end{equation}
where $\tilde{U} \propto r^{-3}$. Here, the distance $r_2$ has been obtained from the relation $\left|C_{3}/r^{3} + C_{6}/r^{6}\right| = \left| C_0\right|$. For $r > r_2$ and $r < L/2$ the interaction is practically constant and given by $\tilde{U} = g^{4}/(\Delta\delta^2) + g^{4}/\delta^3$.\\
There are two special cases for the detuning $\delta$ resulting in the potential forms
\begin{eqnarray}
\label{Eq.C9}
\tilde{U}(r) &=& C_{6}\left(\frac{1}{r^6} - \frac{1}{4R^6} \right) \;\;\;\;\;\;\, \delta = -\frac{\Delta}{2}\\
\label{Eq.C9a}
\tilde{U}(r) &=& C_{6}\left(\frac{1}{r^6} - \frac{1}{2R^3r^3} \right) \;\;\; \delta = -\Delta.
\end{eqnarray}
In Eq.~\ref{Eq.C9} we have a free space van der Waals term followed by a constant all to all interaction at long range while for Eq.~\ref{Eq.C9a} the potential is dominated by the van der Waals term at close proximity and changes at long internuclear distances solely into a dipole-dipole potential form.\\
The mixing of the cavity induced $g_i g_j/\delta$ dipole-dipole interaction and the free space $U_{ij}$, $J_{ij}$ dipole-dipole interaction terms enables these novel dependencies on the internuclear distance.

\section{Ramsey spectroscopy of an ensemble: from the dilute to the dense limit.}
\label{Ramsey}

A method for the investigation of the dynamics of an ensemble of interacting Rydberg atoms for short timescales is time-domain Ramsey interferometry, as recently performed experimentally~\cite{Takei2016direct} and analyzed theoretically~\cite{Sommer2016time}. This method allows for the circumvention of the Rydberg blockade regime and produces a record of the real time evolution of the electronic Rydberg states. For longer timescales and weaker interaction strengths, a frequency-domain Ramsey sequence, as employed in~\cite{Nipper2012Atomic} is also suitable and leads to the same results as found by the former mentioned technique. The Ramsey procedure roughly amounts to transferring population from the ground state $\ket{g}$ into the excited state $|d\rangle$ by a sequence of two time-delayed (delay $\tau$) two-photon excitation pulses allowing interference fringes to be formed. Here, the width of such a pulse $\Delta \omega$ needs to be sufficiently broadband ($\Delta \omega > \tilde{U}(r_{\mathrm{min}})$) to avoid any Rydberg blockade~\cite{Tong2004Local}. For single Rydberg atoms or dilute samples with weak interactions, the periodicity of the fringes reflect the natural evolution of the Rydberg level. For high density samples, as considered experimentally in Ref.~\cite{Takei2016direct} and theoretically analyzed in Ref.~\cite{Sommer2016time}, the fringes are delayed as well as reduced in amplitude as a result of particle-particle interactions.\\
The build-up of correlations is however typically strongly limited by decay and dephasing processes in the system which limit the maximum allowed $\tau$ (for example due to technical limitations to hundreds of $ps$ in the experiment of Ref.~\cite{Takei2016direct}). During such short times, even for high density samples, owing to the rapid falling off of the free space van der Waals coupling with distance, the number of effective atoms participating in the interaction is fairly small~\cite{Takei2016direct}. One could therefore benefit from all-to-all interactions allowing the whole sample to participate in the interactions even for very small $\tau$ (and thus lifting the requirement of having high density ensembles).\\
We therefore proceed to analytically evaluate the characteristics of the Ramsey signal in a case of $N$ intracavity atoms placed within a wavelength and coupled to each other only via the $C_0$ mechanism. According to~\cite{Sommer2016time} the time-domain Ramsey signal is
\begin{eqnarray}
\label{Eq.R0}
P(\tau) = 2p_{g}p_{d}\Re\left\{ 1 + e^{i(\tilde{\omega}_{dg} \tau + \xi)}G(\tau) \right\},
\end{eqnarray}
where $p_{g}$, $p_{d}$ are the population in the ground and excited state, respectively, $\tilde{\omega}_{dg} = \tilde{\omega}_{d}-\omega_{g}$ is the frequency difference between the ground and excited state and $\xi$ is a constant phase resulting from ac-Stark shifts during the pump and probe pulses. The Ramsey signal for a frequency-domain sequence is similar to the expression in Eq.~\ref{Eq.R0} except that $\tilde {\omega}_{dg}$ needs to be exchanged with the detuning $\Delta_{dg} = \omega_{l} - (\tilde{\omega}_{d}-\omega_{g})$, where $\omega_{l}$ is the frequency of the excitation laser.
\begin{figure}[t]
\includegraphics[width=1.0\columnwidth]{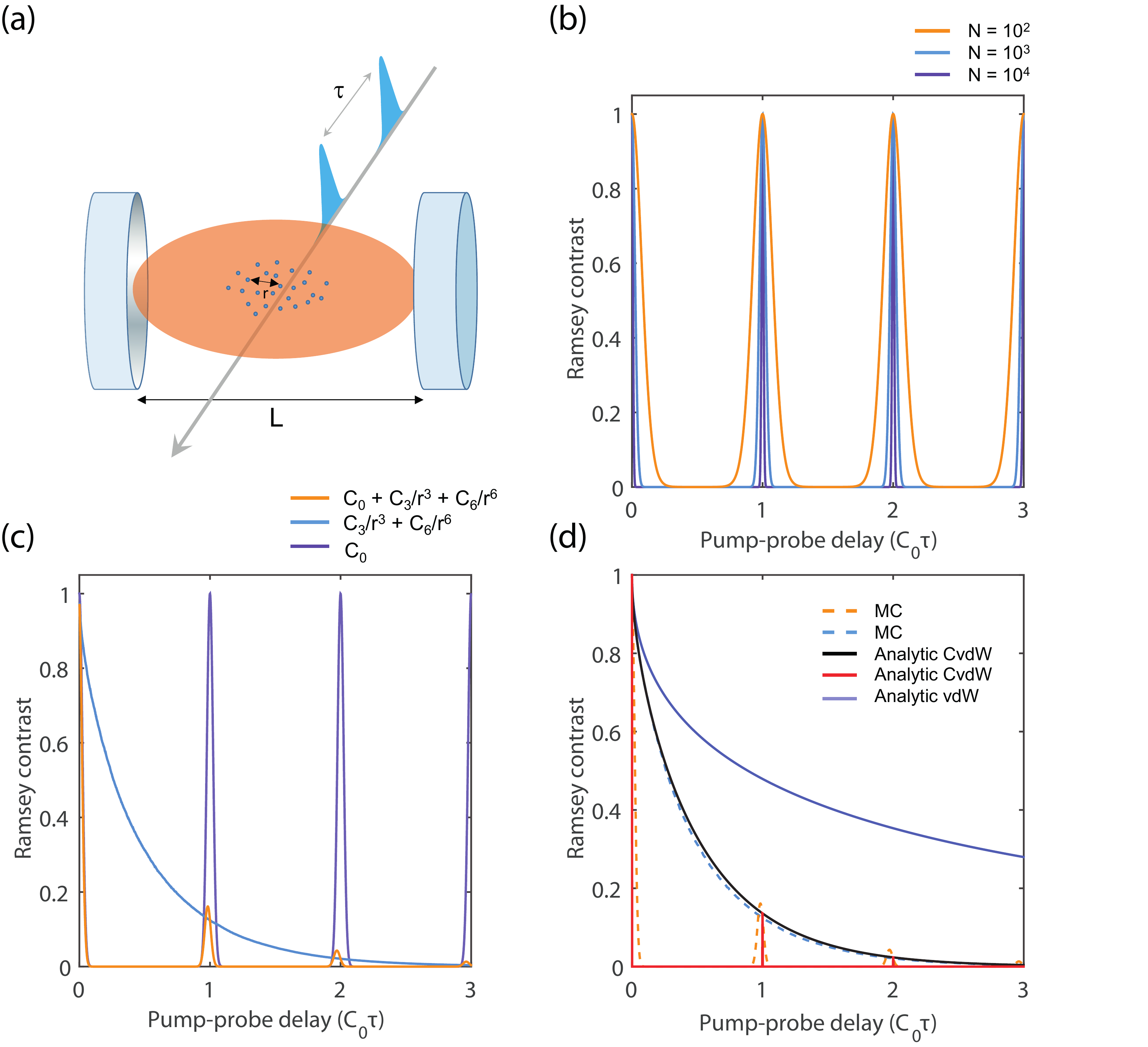}
\caption{\emph{Ramsey interferometry}. (a) Sketch of the Ramsey procedure for an ensemble coupled to the fundamental mode of a microwave cavity. (b) Evolution of the contrast function for all-to-all interactions for $N=100$ (orange line), $N = 1000$ (blue line) and $N = 10000$ (purple line) atoms. (c) Monte-Carlo simulation for $N=1000$ atoms comparing the contrast decay for the full interaction (orange line), intermediated (blue line) and all-to-all (purple line). (d) Comparison of Monte Carlo simulation (dashed lines) to the analytical model (solid lines) in the large particle number limit.  The parameters are $C_{6} = 10C_{0}$, $C_{3} = 10 C_{0}/4$, $n = 0.35\,\mu$m$^{-3}$ and $p_{d} = 5 \%$.}
\label{fig4}
\end{figure}
The interaction modulation of the Ramsey signal \cite{Worm2013Relaxation,Feig2013Dynamical,Hazzard2013Far,Hazzard2013Quantum} is encoded in the term
\begin{eqnarray}
\label{Eq.R1}
G(\tau) &=& \prod_{k\neq j}\left(p_{g} + p_{d}e^{i\tilde{U}_{jk}\tau}\right)\\
\label{Eq.R2}
&\approx& \left( p_{g} + p_{d}e^{iC_{0}\tau}\right)^{N-1} = A^{N-1}e^{i(N-1)\zeta}.
\end{eqnarray}
The quantities of interest experimentally accessible are the contrast $\mu(\tau)=|G(\tau)|$ and phase $\nu(\tau)=-i\ln {(G(\tau)/|G(\tau)|)}$ functions. The contrast can be written as
\begin{equation}
\mu(\tau)=A^{N-1} = \left[p_{g}^{2} + p_{d}^{2} + 2p_{g}p_{d}\cos(C_{0}\tau)\right]^{(N-1)/2},
\end{equation}
while the phase is derived from $\nu(\tau)=(N-1)\zeta$ where
\begin{equation}
\tan(\zeta) = \frac{p_{d}\sin(C_{0}\tau)}{p_{g} + p_{d}\cos(C_{0}\tau)}.
\end{equation}
In the case that $p_{g} = p_{d} = 1/2$ the contrast function becomes $\mu(\tau)= \left|\cos(C_{0}\tau/2)\right|^{N}$. For $N \rightarrow \infty$ this converges to $1$ if $\tau = 2\pi k/C_{0}$, for positive integer $k$ and $0$ otherwise. Figure.~\ref{fig4}b displays the contrast for a finite amount of atoms. The width of the revival features changes $\propto \sqrt{N}$ in the cavity.\\
The very simple expressions obtained above are however strongly altered by the presence of the $C_3$ and $C_6$ terms. For higher densities we perform Monte-Carlo simulations presented in Fig.~\ref{fig4}c. Here, a Monte-Carlo simulation consists of generating random atom locations to form a homogeneous ensemble of atoms of a given density that can be used to calculate the interaction energies $\tilde{U}_{ij}$ for Eq.~\ref{Eq.R1}. On the other hand, we can find an analytic solution for the Ramsey signal at the revival times given by $\tau = 2\pi k/C_{0}$.
The derivation follows a similar route as has been presented in~\cite{Sommer2016time} and is outlined in detail in Appendix B. For a large number of atoms and a locally homogeneous atom distribution of the ensemble, the interaction induced modulation term $G(\tau)$ in Eq.~\ref{Eq.R1} can be approximated by
\begin{eqnarray}
\label{Eq.R3}
G(\tau) &\approx& (p_{g} + p_{d}\gamma(\tau))^{N-1},
\end{eqnarray}
where
\begin{eqnarray}
\gamma(\tau) = \frac{3}{r^{3}_{0} - r^{3}_{\mathrm{B}}}\int^{r_0}_{r_{\mathrm{B}}} dr r^{2} e^{i\tilde{U}(r)\tau}.
\end{eqnarray}
This transition to a continuum description allows us to obtain simple analytical solutions by evaluating $\gamma(\tau)$ in the case of $N \rightarrow \infty$. For simplicity we have also taken $r_{\mathrm{B}} \rightarrow 0$.
\begin{table*}
\begin{tabular}{| l || l | l | l | l | l | l | l | l |}
\hline
State & $\omega_{d}$ (THz) & $\delta$ (GHz) & $\Delta$ (GHz) & $\mu$ $(a_0 e)$ & $g$ (MHz) & $C_{0}$ (MHz) & $C_{3}$ (MHz$\mu$m$^{3}$) & $C_{6}$ (MHz$\mu$m$^{6}$) \\ \hline
$5$D$_{5/2}$ & $2\pi \times 57$ & $14$ & $2.4\times 10^4$ & $10$ & $1.4\times 10^3$ & $1.3$ & $1\times 10^{-3}$ & $4\times 10^{-10}$ \\\hline
$12$D$_{5/2}$ & $2\pi \times 1.7$ & $0.12$ & $31$ & $100$ & $12.3$ & $1.3\times 10^{-2}$ & $0.1$ & $3\times 10^{-3}$\\\hline
$35$D$_{5/2}$ & $2\pi \times 0.12$ & $0.01$ & $1.5$ & $560$ & $0.34$ & $3.8\times 10^{-4}$ & $0.35$ & $63.1$\\\hline
\end{tabular}
\caption{\emph{Numerical estimates.} For $^{87}$Rb the effective potential coefficients are listed for states $5$D$_{5/2}$, $12$D$_{5/2}$ and $35$D$_{5/2}$. Here, the cavity mode volume is set to $V = (\lambda/2)^3$ for each state, respectively. The transition energies and dipole moments have been obtained following~\cite{Gallagher1994Rydberg} and by applying the Numerov method as outlined in~\cite{Blatt1967practical}.}
\label{tab1}
\end{table*}
In the case that $\tau = 2\pi k/C_{0}$ for the general interaction or for the particular case where the detuning is set to cancel the constant all to all interaction we can obtain a finite solution for the amplitude of $G(\tau)$ which is given by
\begin{eqnarray}
\label{Eq.R8}
|G(\tau)| &=& \lim_{N\rightarrow \infty}\left( 1 - \frac{p_d\kappa\tau}{2N}\left(\frac{\pi}{2} + F(\tau) \right)  \right. \\\nonumber
& & -\frac{2p_d\sqrt{\eta\tau}\kappa}{N}\left[\left( \sqrt{\frac{\pi}{8}}- S\left(\sqrt{\frac{\tau}{4\eta}}\right)\right)\cos\left(\frac{\tau}{4\eta}\right) \right. \\\nonumber
& & \left.\left. - \left(\sqrt{\frac{\pi}{8}} - C\left(\sqrt{\frac{\tau}{4\eta}}\right) \right)\sin\left(\frac{\tau}{4\eta}\right) \right]\right)^{N-1}\\
\label{Eq.R9}
&=& e^{- \frac{p_d\kappa\tau}{2}\left(\frac{\pi}{2} + F(\tau) \right)}\\\nonumber
& & \times e^{-2p_d\sqrt{\eta\tau}\kappa\left[\left( \sqrt{\frac{\pi}{8}}- S\left(\sqrt{\frac{\tau}{4\eta}}\right)\right)\cos\left(\frac{\tau}{4\eta}\right) \right]}\\\nonumber
& & \times e^{2p_d\sqrt{\eta\tau}\kappa\left[\left(\sqrt{\frac{\pi}{8}} - C\left(\sqrt{\frac{\tau}{4\eta}}\right) \right)\sin\left(\frac{\tau}{4\eta}\right) \right]},
\end{eqnarray}
where $\eta = C_{6}/C^{2}_{3}$ and the functions $S(x),C(x)$ and $F(\tau)$ are defined in Appendix B. The first exponential term in Eq.~\ref{Eq.R9} governs the contrast degradation at large time delays while the other two exponential terms dominate the decay at an early stage.
The solution is illustrated in Fig.~\ref{fig4}d. Here, good agreement with the results obtained from Monte-Carlo simulations which are depicted by the dashed curves, are found.
In the former case we also know that the amplitude becomes $|G(\tau)| = 0$ when $\tau \neq \frac{2\pi k}{C_{0}}$ which allows us to determine the contrast for all $\tau$ in the general case of the cavity mediated van der Waals interaction.\\
The phase on the other hand shows no finite solution for all $\tau$ and strongly depends on the number of atoms involved in the process. This is due to the trigonometric integral $\mathrm{Ci}(x)$ emerging from the dipole-dipole interaction which diverges for $x\rightarrow 0$. For comparison we present a Ramsey contrast that solely describes the contribution of the free space van der Waals interaction ($\tilde{U}^{F}(r) = C_{6}/r^{6}$) which is expressed by $|G^{F}(\tau)| = e^{-2p_d\sqrt{\frac{\pi}{8}}\kappa\sqrt{\eta \tau}}$ and illustrated as well in Fig.~\ref{fig4}d.\\

\section{Discussions}
The tuning knobs to access the various regimes of the interaction are given by the density of atoms $n$, the detuning between the cavity and the $d \leftrightarrow f$ transition $\delta$, the F\"orster detuning $\Delta$ as well as the cavity frequency $\omega$ and the dipole matrix elements $\mu^{a,b}$ that change with the principal quantum number $\nu$ of an atom. However, simple scaling arguments already indicate an optimal operation regime. Let us assume a Fabry-Perot cavity of length $L$ and waist $w$ such that $V=\pi w^2 L$. Working on a given resonance $\omega_m = 2\pi c/\lambda_m$ with $L=m\lambda_m/2$, the mode volume is expressed as $V=m\pi^2 w^2 c /\omega_m$. The optimization of the ratio $C_0/C_6 \propto \omega^2/V^2 \propto \omega_m^4/m^2$ then obviously requires that one chooses transitions with a high frequency difference. Given that, for high principal quantum numbers the typical difference between $d$ states and neighboring $p,f$ states is small, at the level of $100$ GHz or lower, it is then desired to work with lower levels. We list certain state configurations and detuning conditions in Tab.~\ref{tab1} for the case of $^{87}$Rb atoms as an example that allows for sufficiently strong cavity induced terms. For example, the $5$D$_{5/2}$ state that couples strongly to the energetically higher $4$F$_{7/2}$ state and the energetically lower $6$P$_{3/2}$ state can have a strong cavity induced constant van der Waals interaction around $1$ MHz over an internuclear distance range of $\lambda/4 \approx 1.3\,\mu$m to $100\,$nm where the cavity induced dipole-dipole term starts to dominate. For atoms trapped in an optical lattice with a lattice constant of $250\,$nm up to $\sim 130$ atoms can be coupled simultaneously via the distance independent interaction. An example where the cavity induced dipole-dipole interaction dominates the dynamics is given for $12$D$_{5/2}$ which couples strongly to $13$P$_{3/2}$ and $11$F$_{7/2}$. This is true for a range extending from $330\,$nm to $2\,\mu$m, while for smaller internuclear distances the free space van der Waals and for larger distances the constant all to all interaction (ranging to $\lambda/4 = 44\,\mu$m) govern the dynamics, respectively. By choosing the detuning appropriately we can find a minimum as presented in Fig.~\ref{fig3}a,c forming a binding potential for two dimensional arrangements of atoms in this region. For much higher principal quantum numbers $\nu$ the cavity induced interaction terms become small in comparison to the increasing free space van der Waals interaction as it has been presented for $35$D$_{5/2}$ coupling to $37$P$_{3/2}$ and $33$F$_{7/2}$ in Tab.~\ref{tab1}. Nevertheless, for $35$D$_{5/2}$ and sufficiently low densities with internuclear distances ranging from $7\,\mu$m to $\lambda/4 = 630\,\mu$m the cavity induced terms dominate with sub MHz strength. Finally, a full experimental feasibility study will have to account for a plethora of experimental detrimental effects among which, for example, are magnetic/electric stray fields. These will modify the natural frequency of the atoms. However, as cloud sizes are quite reduced (order of microns), it is justified to assume that all atoms will have the same shift at the same time. Moreover, experiments as in Ref.~\cite{Takei2016direct} are performed on the picosecond to nanosecond timescale. During such short durations, one would expect that magnetic/electric field fluctuations (usually in the kHz to MHz regime) will have a negligible impact on the signature of the temporal dynamics.

\section{Conclusions}
We have shown that manipulating the density of modes of the electromagnetic vacuum field by means of a microwave cavity can strongly alter the van der Waals interaction between Rydberg atoms in an ensemble. The main result indicates the possibility of switching between nearest neighbor to all-to-all interaction regimes. We have furthermore analyzed a particular situation involving a standard Fabry-Perot microwave cavity and concluded that experimental feasibility requires the use of Rydberg manifolds with low principal quantum numbers.

\section{Acknowledgements}
We acknowledge financial support from the Max Planck Society.

\bibliography{Rydbib}

\newpage

\section{Appendix A: Reduced Hamiltonian in the two excitation basis}
\label{App:1}
In the absence of any decay mechanism the dynamics of the system is fully described by the Hamiltonian matrix
\begin{eqnarray}
\label{Eq.C1}
H &=& \left(\begin{smallmatrix} 2\omega_{d}-\delta & J & \sqrt{2}g^{a}_{1} & g^{b}_{1} & 0 & g^{a}_{2} \\
J & 2\omega_{d}-\delta & \sqrt{2}g^{a}_{2} & 0 & g^{b}_{2} & g^{a}_{1} \\
\sqrt{2}g^{a}_{1} & \sqrt{2}g^{a}_{2} & 2\omega_{d}-2\delta & 0 & 0 & 0 \\ g^{b}_{1} & 0 & 0 & 2\omega_{d}-\Delta & 0 & U \\ 0 & g^{b}_{2} & 0 & 0 & 2\omega_{d}-\Delta & U\\ g^{a}_{2} & g^{a}_{1} & 0 & U & U & 2\omega_{d} \end{smallmatrix} \right),
\end{eqnarray}
where $\delta = \omega_{d}-\omega$ and $\Delta = 2\omega_{d}-\omega_{p}$.
The expression in Eq.~\ref{Eq.C1} is quite general but can be simplified for smaller distances $r$ between the atoms where the coupling strengths $g^{a,b}$ become equivalent at each atom site. Here we obtain the reduced matrix
\begin{eqnarray}
\label{Eq.C2}
H &=& \left(\begin{array}{cccc} 2\omega_{d}-\delta+J & 2g^{a} & g^{b} & \sqrt{2}g^{a} \\
2g^{a} & 2\omega_{d}-2\delta & 0 & 0 \\
g^{b} & 0 & 2\omega_{d}-\Delta & \sqrt{2}U \\
\sqrt{2}g^{a} & 0 & \sqrt{2}U & 2\omega_{d} \end{array} \right)
\end{eqnarray}
with respect to the basis states $1/\sqrt{2}(\ket{df1} + \ket {fd1})$, $\ket {ff2}$, $1/\sqrt{2}(\ket {pf0} + \ket {fp0})$, $\ket {dd0}$.\\

\section{Appendix B: Continuum description of the Ramsey signal}
\label{App:2}
With $\tilde{U}(r) = C_{0} + C_{3}/r^{3} + C_{6}/r^{6}$ we can reformulate
\begin{eqnarray}
\label{Eq.R4}
\gamma(\tau) &=& \frac{3}{r^{3}_{0} - r^{3}_{\mathrm{B}}}\int^{r_0}_{r_{\mathrm{B}}} dr r^{2}e^{i\left(C_{0} + \frac{C_{3}}{r^{3}} + \frac{C_{6}}{r^{6}} \right)\tau}\\
\label{Eq.R5}
&=& \frac{\omega_{0}\omega_{\mathrm{B}}}{(\omega_{\mathrm{B}}-\omega_{0})}e^{iC_{0}\tau}\int^{\omega_{\mathrm{B}}}_{\omega_{0}}d\omega \frac{1}{\omega^{2}}e^{i(\omega + \eta \omega^{2})\tau},\nonumber
\end{eqnarray}
where $\omega = C_{3}/r^{3}$ and
\begin{equation}
\eta = \frac{\Delta\delta^{2}}{8\left( g^{a}g^{b}\left(1+\frac{\Delta}{2\delta}\right)\right)^{2}}
\end{equation}
resulting from $C_{6} = \eta C^{2}_{3}$. Integration by parts and various substitutions lead to
\begin{eqnarray}
\label{Eq.R6}
\gamma(\tau) &=& e^{iC_{0}\tau}\left \{ \left(\frac{\omega_{\mathrm{B}}e^{i(\omega_0 + \eta\omega^2_0)\tau}-\omega_0 e^{i(\omega_{\mathrm{B}} + \eta\omega^2_{\mathrm{B}})\tau}}{(\omega_{\mathrm{B}}-\omega_{0})} \right) \right. \\\nonumber
& & + \frac{\omega_{0}\omega_{\mathrm{B}}}{(\omega_{\mathrm{B}}-\omega_{0})}\left( 2\sqrt{\eta\tau}e^{-\frac{\tau}{4\eta}}\left(i\left[C(\hat{\omega}_{\mathrm{B}}\sqrt{\tau})-C(\hat{\omega}_{0}\sqrt{\tau}) \right]\right. \right. \\\nonumber
& & \left. - \left[S(\hat{\omega}_{\mathrm{B}}\sqrt{\tau})-S(\hat{\omega}_{0}\sqrt{\tau})\right] \right) \\\nonumber
& & -\frac{\tau}{2}\left[\mathrm{Si}((\omega^{2}_{\mathrm{B}}\eta + \omega_{\mathrm{B}})\tau) + \mathrm{Si}_{\mathrm{M},\frac{4\eta}{\tau}}((\omega^{2}_{\mathrm{B}}\eta + \omega_{\mathrm{B}})\tau) \right. \\\nonumber
& & \left. - \mathrm{Si}((\omega^{2}_{0}\eta + \omega_{0})\tau) - \mathrm{Si}_{\mathrm{M},\frac{4\eta}{\tau}}((\omega^{2}_{0}\eta + \omega_{0})\tau) \right] \\\nonumber
& & +\frac{i\tau}{2}\left[\mathrm{Ci}((\omega^{2}_{\mathrm{B}}\eta + \omega_{\mathrm{B}})\tau) + \mathrm{Ci}_{\mathrm{M},\frac{4\eta}{\tau}}((\omega^{2}_{\mathrm{B}}\eta + \omega_{\mathrm{B}})\tau)  \right. \\\nonumber
& &\left. \left. \left. - \mathrm{Ci}((\omega^{2}_{0}\eta + \omega_{0})\tau) - \mathrm{Ci}_{\mathrm{M},\frac{4\eta}{\tau}}((\omega^{2}_{0}\eta + \omega_{0})\tau)\right] \right) \right\},
\end{eqnarray}
where $\hat{\omega} = \left(\omega\sqrt{\eta} + 1/2\sqrt{\eta} \right)$, $S(x) = \int^{x}_{0} \sin(t^2)dt$, $C(x) = \int^{x}_{0} \cos(t^2)dt$ are Fresnel integrals, $\mathrm{Si}(x) = \int^{x}_{0} \frac{\sin(t)}{t}dt$, $\mathrm{Ci}(x) = -\int^{\infty}_{x} \frac{\cos(t)}{t}dt$ are trigonometric integrals and we define $\mathrm{Si}_{\mathrm{M},\beta}(x) = \int^{x}_{0} \frac{1}{\sqrt{\beta t+1}}\frac{\sin(t)}{t}dt$, $\mathrm{Ci}_{\mathrm{M},\beta}(x) = -\int^{\infty}_{x} \frac{1}{\sqrt{\beta t+1}}\frac{\cos(t)}{t}dt$ as modified trigonometric integrals that converge for $\beta \rightarrow 0$ against the standard trigonometric integrals and vanish for $\beta \rightarrow \infty$.\\

Using the relation $\omega_0 = 4\pi n C_{3}/(3N) = \kappa/N$, where $n$ is the density of atoms in the spherical volume and by employing the conditions $N \rightarrow \infty$, $\omega_{\mathrm{B}} \rightarrow \infty$, Eq.~\ref{Eq.R6} can be simplified to
\begin{eqnarray}
\label{Eq.R7}
\gamma(\tau) &=& e^{iC_{0}\tau}\left \{ 1 - \frac{\kappa\tau}{2N}\left(\frac{\pi}{2} + F(\tau) \right)  \right. \\\nonumber
& & + i\frac{\kappa\tau}{N}\left(1-\frac{1}{2}\mathrm{Ci}\left(\frac{\kappa\tau}{N}\right) - \frac{1}{2}\mathrm{Ci}_{\mathrm{M},\frac{4\eta}{\tau}}\left(\frac{\kappa\tau}{N}\right) \right) \\\nonumber
& & -\frac{2\sqrt{\eta\tau}\kappa}{N}\left[\left( \sqrt{\frac{\pi}{8}}- S\left(\sqrt{\frac{\tau}{4\eta}}\right)\right)\cos\left(\frac{\tau}{4\eta}\right) \right. \\\nonumber
& & \left. - \left(\sqrt{\frac{\pi}{8}} - C\left(\sqrt{\frac{\tau}{4\eta}}\right) \right)\sin\left(\frac{\tau}{4\eta}\right) \right] \\\nonumber
& & +i\frac{2\sqrt{\eta\tau}\kappa}{N}\left[\left( \sqrt{\frac{\pi}{8}}- C\left(\sqrt{\frac{\tau}{4\eta}}\right)\right)\cos\left(\frac{\tau}{4\eta}\right) \right. \\\nonumber
& & \left.\left. - \left(\sqrt{\frac{\pi}{8}} - S\left(\sqrt{\frac{\tau}{4\eta}}\right) \right)\sin\left(\frac{\tau}{4\eta}\right) \right]\right\},
\end{eqnarray}
where $F(\tau) = \mathrm{Si}_{\mathrm{M},\frac{4\eta}{\tau}}(\infty)$ is a $N$-independent monotonic function that increases from $F(0) = 0$ to $\lim_{\tau \rightarrow \infty} F(\tau) = \frac{\pi}{2}$.\\

\end{document}